\documentstyle[aps,prd]{revtex}

\newcommand{\be}{\begin{equation}}
\newcommand{\ee}{\end{equation}}
\newcommand{\bea}{\begin{eqnarray}}
\newcommand{\eea}{\end{eqnarray}}
\newcommand{\sch}{Schwarzschild }
\newcommand{\r}{\eta}

\begin{document}

\title{Cauchy-characteristic Evolution of Einstein-Klein-Gordon
Systems: The Black Hole Regime}

\author{Philippos Papadopoulos and Pablo Laguna}
\address{Department of Astronomy \& Astrophysics and\\
         Center for Gravitational Physics \& Geometry\\
         Penn State University, University Park, PA 16802}

\maketitle

\begin{abstract}

The Cauchy+characteristic matching (CCM) problem for the scalar wave
equation is investigated in the background geometry of a \sch black
hole. Previously reported work developed the CCM framework for the
coupled Einstein-Klein-Gordon system of equations, assuming a regular
center of symmetry. Here, the time evolution after the formation of a
black hole is pursued, using a CCM formulation of the governing
equations perturbed around the \sch background. An extension of the
matching scheme allows for arbitrary matching boundary motion across
the coordinate grid.  As a proof of concept, the late time behavior of
the dynamics of the scalar field is explored. The power-law tails in
both the time-like and null infinity limits are verified.

\end{abstract}

\section{Introduction}

The evolution of black hole binaries, after the initial inspiral phase
and into the late time ring-down stage, requires the numerical
solution of the full set of Einstein equations. A major scientific
effort, the Binary Black Hole Grand Challenge Collaboration~\cite{GC},
is currently underway for the study of this problem. An important
aspect of this investigation is the condition to be imposed at the
outer boundary of the computational domain.  Assuming a physically
correct prescription of initial data on a Cauchy surface, the
evolution is unambiguous in the domain of dependence of the Cauchy
surface; problems arise when the space region in which the equations
are to be evolved is of fixed finite size, as it is necessarily the
case in numerical simulations. The practical relevance of this issue
increases as longer evolution times are achieved and strong,
non-symmetric field configurations are being probed.

At present, a research program is unfolding
\cite{GC,CCM,3Dccm,ClarkeI,ClarkeII,ClarkeIII}, in which the problem
of radiative outer boundary conditions is dealt with at its heart, by
extending the integration domain to null infinity. The method combines
a characteristic initial value problem, with a regular Cauchy initial
value problem. The characteristic initial value problem is specified
on null cones, emerging from the world-tube boundary of the Cauchy
surface. In previous work by G\'{o}mez et. al. \cite{CCM} (henceforth
GLPW), the CCM framework was applied to a system widely used as a
model for more complicated relativistic systems, namely the coupled
Einstein-Klein-Gordon set of equations in spherical symmetry. At the
core of the matching strategy, continuity conditions on metric,
extrinsic curvature and scalar field variables ensure smoothness
across the matching interface (which is assumed to be at fixed
spherical surface of area $4 \pi R^{2}$). These conditions provide the
practical ``handle'' for transforming geometric and field information
between the two different space-time foliations.  Accuracy tests and
comparison of CCM evolutions with reference solutions performed in
GLPW, show a remarkably transparent propagation of information across
the matching interface, even for strongly non-linear waves.

The choice of gauge in GLPW did not permit the integration of the
equations beyond the formation of a black hole, which is the
generic conclusion of strong field evolutions. Even so, in a
collapse scenario where the newly formed black hole has a mass $M_{BH}
> 2 R$, the exterior characteristic code allows for the computation of
all relevant signal at null infinity, although the computation
stops at a finite coordinate time \cite{1-dscalar,Null-cone}.  In
realistic three dimensional computations though, especially those
planned for the computation of the binary black hole inspiral, the
matching radius $R$ is expected to be at least larger than
$M_{BH}$. In such a case, the CCM code developed in GLPW halts during
the formation of a black hole, and the signal at null infinity is
obtained only up to some finite asymptotic (Bondi) time.

In this paper we focus on this specific aspect of the CCM for the
coupled Einstein-Klein-Gordon system, namely the accurate long-term
evolution {\em after} a black hole has formed. To this end, it is
assumed that a \sch black hole of mass $M_{BH}$ is present in the
space-time, and the remaining scalar field stress-energy does not
further modify the background space-time. This is a situation
resembling the later stages of the black hole collision process, or,
indeed, an arbitrary stage, provided the matching radius is
chosen as $R >> M_{BH}$.

The fixed curved background crystallizes a basic feature of the
colliding black hole space-times, namely the relativistic potential
encompassing the dynamic inner region.  The effects of the potential
on the propagation of waves are well known \cite{DeWitt}, and can be
probed analytically in a number of different ways
\cite{Price,Schmidt,Suen}. Numerically, the detection of the late
time features can be a challenging effort for multi-dimensional
algorithms, as it requires a significant dynamical
range. Nevertheless, the full dynamical calculations of black hole
collisions must continue into the quasi-normal ringing and, possibly,
the tail regime to achieve a complete mapping of the anticipated
signal.

The presence of late time power-law tails has been extensively
verified numerically in one-dimensional problems using both Cauchy
\cite{Suen} and null \cite{Price} initial value formulations in fixed
geometric backgrounds.  The existence of tails has also been
investigated in dynamical self-gravitating systems
\cite{Schmidt,Price2} and rotating space-times \cite{krivan}.  In this
paper the tail behavior in the \sch potential is investigated using
the CCM approach. This serves both as a test of the long term accuracy
of the method, and, importantly, as an illustration of the intrinsic
economy and versatility of the CCM, which adds significant dynamic 
range to the basic Cauchy evolution
algorithm.

The paper is organized as follows: In section \ref{sec:\sch}, the CCM
for scalar waves in a \sch background is presented, as well as the
coordinate choices for both the space-like and null region.  Section
\ref{sec:matching} describes the numerical techniques involved in
performing the matching, in particular, the extensions to the
interpolation procedures employed in GLPW.  Section \ref{sec:results}
reports tests of the CCM algorithm and results.  The stability and
accuracy of the curved background evolution are examined using energy
conservation and Cauchy convergence tests. The case of a moving
matching boundary is illustrated and tested. Lastly, the focus is
turned on the propagation phenomena associated with large evolution
times. The efficiency of the scheme is demonstrated in the calculation
of late-time power law tails at both null and time-like infinity with
moderate resolution requirements.  Established analytic results are
verified readily.

\section{The CCM in a Schwarzschild Space-time}
\label{sec:\sch}

Following the notation of GLPW, the spacetime domains covered by the
Cauchy and characteristic foliations are denoted $M^{-}$ and $M^{+}$,
respectively.  Coordinate systems layed out in the two domains are
assumed to overlap. In the overlap region, a time-like world-tube
serves as the outer boundary of the interior spacelike 
hypersurfaces and the
inner boundary of the exterior null hypersurfaces.

A coordinate system $(t,\r)$ is adopted in
region $M^{-}$, in which the line element is written as
\be
ds^2 = \alpha(\r)^2 ( -dt^2 + d\r^2)  + r(\r)^2 d\Omega^2 \, ,
\label{eq:lineS}
\ee
where $r$ is an area scalar, $t$ defines a slicing in which the
metric is time independent, $ \alpha^2 = 1 - 2 M/r$ and
$d\r / dr = 1/\alpha^{2}$ defines a  ``tortoise'' coordinate $\r$.

The matching world-tube is assumed to be located at $\r_{+}$.

For $r \geq r(\r_{+})$ a coordinate system $(u,x)$ is adopted, where
the compactified radial coordinate $x = r / (1 + r)$, and the retarded
time $u = t - \r$ are employed. The line element in the $M^{+}$ region
takes the corresponding ``null'' form
\be
ds^2 = - \alpha(x)^2 du^2 - 2 (1-x)^2 du dx + r(x)^2
d\Omega^2 \, .
\label{eq:lineN}
\ee

The wave equation $\Box \phi \equiv \nabla_\mu\nabla^\mu \phi = 0$ in
the background metric described by either (\ref{eq:lineS}) or
(\ref{eq:lineN}) is separable and reduces to $1+1$
partial differential equations.  
In $M^{-}$, the wave equation takes the form
\be
g_{,tt} - g_{,\r\r}  =  (1 - \frac{2M}{r}) V(r) g \, ,
\label{eq:schwS} 
\ee
where
\be
V(r) = -\frac{2M}{r^3} + \frac{l(l+1)}{r^2} \, ,
\label{eq:schpot}
\ee
and $g = r \phi $.

In $M^{+}$, the wave equation is converted into an integral
identity. The procedure is an application of the method outlined in
\cite{jcomp} and we do not elaborate here further.  A surface
integration over a bounded region $C$, defined by $u_{1} < u < u_{2}$,
$v_{1} < v < v_{2}$, i.e., it by two outgoing null geodesics
$(u=u_{1}, u=u_{2})$ and two incoming null geodesics $(v=v_{1},
v=v_{2})$ leads to the identity
\be
     g_{u_{2} v_{2}} = g_{u_{2} v_{1}}
                     + g_{u_{1} v_{2}}
                     - g_{u_{1} v_{1}}
     - {1 \over 2} \int_C V g \, du dr \, , 
\label{eq:schwN}
\ee
which relates the values of the scalar field at the vertices of C with
the integral of the potential term.  An important fact deserves
mentioning here, the radial variable $x$ introduces a
``compactification'' of the radial variable $r$ in the null patch
$M^{+}$. This transformation locates the future null infinity of the
space-time at a finite coordinate distance. The value of the field at
infinity is hence obtained as the last step of the radial integration
of (\ref{eq:schwN}).

The initial value problem in the $M^{-}$ domain is completed with the
specification of the inner and outer boundary conditions. The inner
boundary condition is imposed at a point $\r_{-}$ sufficiently close
to the horizon. For the purposes of this investigation a value $\r_{-}
= -25$ is sufficient.  The boundary condition imposed at $\r_{-}$ is
that of a purely in-going wave. The exponential fall-off of the \sch
effective potential $V$, as $\r \rightarrow - \infty $, ensures that
purely in-going wave solutions converge to the exact solution
exponentially. This situation is in sharp contrast with the power law
fall-off in the outer region.  The outer boundary condition is the
subject of section \ref{sec:matching}. Algorithmically, the matching
condition at the outer boundary completes the evolution of the inner
region $M^{-}$ by supplying field information at $\r_{+}$, the
coordinate location of the outer boundary.

The next step is to produce discrete approximations of the wave
equations (\ref{eq:schwS}) and (\ref{eq:schwN}).  In the coordinate
system $(t,\r)$ and with the definition $p \equiv g_{,t}$, the wave
equation (\ref{eq:schwS}) is written as a first order system,
\bea
g_{,t} & = & p \, , \label{eq:10} \\
p_{,t} & = & g_{,\r\r} - \biggl(1-\frac{2M}{r}\biggr) V g.
\label{eq:11}
\eea
The tortoise coordinate $\r$ is given in the case of the
\sch metric by
\be
\r = r + 2 M \ln(\frac{r}{2M} - 1).
\label{eq:tortoise}
\ee
The inverse function $r(\r)$  is formally
given in terms of the principal part of the Lambert function $W(y)$
\be
r(\r) = 2 M \lbrace 1 + W(e^{\r/2M - 1})\rbrace,
\ee
where $W(y)$ solves the equation
\be
W(y) e^{W(y)} = y.
\ee

A direct numerical iteration of equation (\ref{eq:tortoise}) using the
Newton-Raphson method provides accurate values for the inverse
function at any required point $r$.  The temporal discretization of
the wave equation (\ref{eq:10}) and (\ref{eq:11}) in $M^{-}$ employs a
modified staggered leapfrog scheme. The field variable $g(t,\r)$ is
approximated by values $g^{n}_{i}$ at the discrete time levels $t^n$
and spatial points $\r_{i}$, with $i$ ranging between $0$ (at
$\r_{-}$) and $N$ (at $\r_{+}$).  The momentum variable $p(t,\r)$ is
approximated at the staggered time levels $t^{n+1/2}$.  The discrete
time-step and grid spacing are denoted $\Delta t$ and $\Delta \r$
respectively.  The spatial discretization of the second derivative
$g_{\r\r}$ is of fourth order accuracy for $1<i<N-1$ and of second
order accuracy otherwise. The condition for stable evolution of smooth
initial data, i.e., the suppression of non-physical degrees of
freedom, is given by the standard stability condition $ \Delta t < K
\Delta \r $, where $K = 0.5$ for marginal stability.

In the $M^{+}$ region too, following the finite difference paradigm,
a spatio-temporal grid $(u^m, x_i)$ is introduced, with discrete
time-steps and length-scales of $\Delta u$ and $\Delta x$
respectively.  The field variable $g$ is approximated by a discrete
set of values $g^m_i$ at those points.  In order to obtain a discrete
version of (\ref{eq:schwN}), the region of integration $C$ must be
suitably placed on the grid. The null parallelogram $C$ is placed with
the inward characteristics centered on grid points $x_i$.  The
equation of the in-going characteristics of the metric is
\begin{equation}
\frac{dx}{du} = - \frac{1}{2} \biggl(1 - \frac{2M}{r}\biggr) (1-x)^2,
\label{eq:ingeo}
\end{equation}
where $r = x/(1-x)$.
The integration of the null geodesic (\ref{eq:ingeo}) provides the
coordinates of the vertices.  The values of the field at the vertices
are to be interpolated from the grid values $g^m_{k}, g^{m+1}_{k}$
where $k$ spans the range $[i-2,i+1]$,  $i$ being the index of the new
point under evaluation. (See \cite{jcomp}).

The integral in the right hand side of equation~(\ref{eq:schwN}) 
is approximated as
\be
 \int_C V g \, du dr  \approx - g_\star \,
\left[ l (l+1) \int_C \frac{du dr}{r^2} +
      2 M \int_C \frac{du dr}{r^3} \right] \, , 
\label{eq:12}
\ee
where $g_\star \equiv g(r_\star)$.
The point $r_\star$ is at the center of the null parallelogram, hence,
$g_\star$ can be obtained at $r_\star$ by averaging
$g_\star = (g_{u_{1} v_{2}} + g_{u_{2} v_{1}})/2 $.
The remaining integrals can be evaluated analytically, yielding
\bea
 \int_C \frac{du dr}{r^2} & = & 2
\log{ \frac{(r_{Q} - 2 M) (r_{R} - 2 M)}{(r_P - 2 M)(r_S - 2M)}  } \, ,\\
 \int_C \frac{du dr}{r^3} & = &
\frac{1}{2M} \log{ \frac{(1-2M/r_Q)(1-2M/r_R)}{(1-2M/r_P)(1-2M/r_S)} },
\eea
where $P,Q,R,S$ denote the points
$(u_2,v_1),(u_2,v_2),(u_1,v_1),(u_1,v_2)$ respectively.
Substitution of these integrals into (\ref{eq:schwN}) yields the algorithm
for updating the field $g$ in the $M^{+}$ sector.

The domain of stability for this updating algorithm is
dictated by the CFL condition, which requires that
\be
\Delta u < 2 \frac{\Delta x}{(1 - x_{m})^2} \frac{1}{1 + 2 M - 2 M /x_{m}},
\ee
where $x_{m}$ is the location of the innermost point where the
method is applied.

\section{The Matching interface}\label{sec:matching}

A smooth matching interface between the two coordinate domains
requires that the scalar field, and any geometric quantities
constructed from it, transform properly under the coordinate
transformation connecting the two spacetime patches.  Again, following
the notation and exposition of GLPW, scalar quantities evaluated at the
interface must be invariant: \bea
[\phi]^{-} & = & [\phi]^{+}  \, ,
\label{eq:mphi} \\
\lbrack{\bf k} \cdot \nabla \phi\rbrack^{-} 
 & = &
\lbrack{\bf k} \cdot \nabla \phi\rbrack^{+} \, ,
\label{eq:mphir}
\eea
where the ${\bf k}$ is an arbitrary vector at the world-tube boundary
and $\nabla$ is the metric connection.  The first condition is
sufficient for the construction of a matching interface, yet practical
matching schemes can be based on combinations of (\ref{eq:mphi}) with
various forms of (\ref{eq:mphir}).

The strategy for the matching must take into account technical issues
arising from the adoption of a finite difference framework. The
numerical solution to the initial value problem replaces the continuum
domain with a finite set of points where the solution is sought to be
approximated. The boundary on which conditions like (\ref{eq:mphi})
and (\ref{eq:mphir}) are true, need not necessarily lie on grid
points, and hence the enforcement of those conditions calls for a
systematic procedure for interpolating information between the two
domains. The numerical interpolation scheme developed here, in
addition to enforcing the matching conditions, also provides the
flexibility of a {\em moving}, in coordinates, boundary of the two
domains.  Such an arrangement may be of importance in future CCM
investigations as it brings to the CCM method a certain degree of {\em
adaptivity}.  In this work though, the moving boundary case is studied
only as a proof of principle.

The collection of $M^{-}$ grid points that cannot be evolved using the
evolution equation are denoted as {\em boundary points}. In the
present one-dimensional problem, and for second order spatial
discretization near the boundary, there is only a single time
sequence of such points, namely $(t^n,\r^N)$.  At boundary points, 
one may generally need information
about both the fields and their first derivatives in time and
space. The basic technique of the matching procedure is to generate
this information by transforming field values from the $M^{+}$ domain:
The algorithm starts with {\em cubic} interpolations along the radial
null $x$ directions on each of the retarded time levels
$u^{n-1},u^{n-2},...$. The radial locations at which interpolation is
performed are at the cross-sections of the $t^n$ level with each of
the $u^k$ levels.  This computation provides an $O(\Delta^4)$ accurate
approximation of the field values on the extension of the $t^n$ level
outside the boundary.  Next, a {\em cubic spline}
interpolation combines values from the regular $M^{-}$ grid at the
$t^n$ level, with the newly interpolated values outside the boundary.
Thus, an $O(\Delta^4)$ accurate approximation of the field can be
obtained anywhere in the neighborhood of the boundary. In the problem
at hand, the interpolant is used to provide the value for the boundary
point of the $t^n$ level and the first point of the $u^n$ level.  The
use of spline interpolation enforces smooth first order radial
derivatives of the field. Thus, the interpolant satisfies identically
both (\ref{eq:mphi}) and (\ref{eq:mphir}) conditions.

The practical implementation of the interpolation procedure asks for a
minor departure from pure evolution algorithms, in that field values
in the neighborhood of the boundary are stored for a sufficient number
of iterations.  In the present context, i.e., a fixed static geometric
background, more minimalistic schemes are possible.  Nevertheless, the
interpolation procedure has been constructed in a way that will allow
the handling of some more complicated matching scenarios.

\section{Tests and Results}\label{sec:results}

The stability and convergence characteristics of the CCM algorithm
described in the previous sections rely in a significant way on the
corresponding properties of the individual components, namely the
Cauchy and characteristic evolutions. The standard discretization
scheme used in the space-like $M^{-}$ region assures that, within the
stability constraints of the explicit scheme used, the numerical
evolution in $M^{-}$ is stable. The characteristic algorithm, is also
an application of a carefully calibrated, conditionally stable, scheme
\cite{jcomp}.  The matching of the two evolution algorithms
introduces, on the other hand, a new approximate operation that must
be validated accordingly. In GLPW, extensive testing of the algorithm
established its stability in the presence of local non-linear wave
propagation. The tests presented here suggest further, that in the
presence of a black hole, for a variety of initial data and for long
evolution times, no growing modes become manifest.

Energy conservation and Cauchy convergence are two standard tools for
assessing the accuracy characteristics of an evolution algorithm. In
the case under consideration, the conservation law can be used to
assess the accuracy with which energy is transported across the
matching surface. The total energy content $E$ of a space-like,
hypersurface in $M^{-}$ labeled by $t$ is given by
\begin{equation}
E(t) = \frac{1}{2} \int_{\r_{-}}^{\r_{+}}
(g_{,t}^2 + g_{,\r}^2 + V(\r) g^2) d\r \, .
\end{equation}
The change of total energy in a time interval $t_1-t_0$ 
is equal to the time integral of the flux across the boundaries; that is,
\begin{equation}
E(t_1) - E(t_0) = F(t_1,t_0) = 
\int_{t_{0}}^{t_{1}} g_{,t} g_{,\r} dt \biggl|_{\r_{-}} - 
\int_{t_{0}}^{t_{1}} g_{,t} g_{,\r} dt \biggr|_{\r_{+}}    .
\end{equation}
This identity allows a robust check on the consistency and accuracy of
the matching. In Fig.~\ref{fig:energy-con}, the second order accuracy
of energy conservation is displayed. The plot is produced by grid
sizes of $(200 \times 2^k + 200 \times 2^k)$ points, where
$k=1,2,3,4$.  (I.e., the coarsest grid resolution for both regions has
been taken the same).  The initial data are an outgoing polynomial
pulse, with compact support in $-4 < \r < -1$. The matching radius is
taken at $\r = 2$, inside the peak of the \sch potential well, in
order to check energy transmission in the most sensitive region.  The
upper plot demonstrates the time development of the energy content
$E(t)$ of $M^{-}$, along with $F(t,t_0)$, the accumulated radiated
power through the interface. The lower plot demonstrates the
convergence of the quantity $\delta = E(t_1)-E(t_0)-F(t_1,t_0)$ to
zero, as a function of grid size, for $t_0 =0$, $t_1=6$. The diamond
points are numerical results, while the solid line is a least square
fit indicating a slope of $1.995$.

As mentioned above, the interpolation procedure at the boundary allows
for considerable freedom of boundary motion. At present, it is not
clear if such freedom is going to be necessary in realistic
computations. Nevertheless, it is interesting and indicative of the
robustness of the method to study the behavior of the mixed initial
value problem in cases where the boundary of the domains is time
dependent. A simple example of a moving boundary evolution in the
presence of an oscillating source is demonstrated in
Fig.~\ref{fig:moving} (in this case M=0).  The original position of the
boundary is at $r=50$ and it subsequently moves inwards, with a
constant speed.  The initial data are taken to be zero in both domains.
The region to the left of the boundary displays the field in the
$(t,r)$ coordinate, while the region to the right of the boundary
shows retarded field evolution. A number of source oscillations are
visible near the origin, along with outgoing shells of radiation that
are intercepted by the ingoing boundary and emmited to null infinity.
The speed of the boundary is of the order of the local wave speed and
does not appear to be restricted in the subluminal range.  Long
evolutions, with oscillating boundaries and constant supply of energy,
indicate that the boundary motion does not introduce any significant
error besides the nominal interpolation error.

Another test of the algorithm focuses on generic aspects of the wave
propagation in a curved background.  The qualitative features of the
evolution of compact support data in the \sch background are well
established in the literature.  In flat three dimensional space, the
wave equation satisfies the Huygens principle, i.e., signals with
sharp wavefront remain sharp. In a curved background, the
corresponding partial differential equation does not satisfy this
principle; this feature has quite dramatic impact on the late time
evolution of initial data with compact support.  The evolution in the
\sch background is predominantly affected by the effective
potential. After the initial pulse propagates away from the potential
region, radiation energy is being released in the form of damped
oscillations that decay exponentially.  During the decay of those
``quasi-normal'' mode oscillations, another feature of the \sch
potential emerges; the long range of the potential allows the
radiation to backscatter and reach the observer at arbitrarily late
times, giving rise to power-law tails. The accurate description of the
different aspects of the evolution is given in terms of the structure
of the Green's function for a one dimensional wave equation with a
potential term. This function is seen to include three
contributions, the flat wave propagator, poles in the complex
frequency domain that are responsible for the damped oscillations and
a branch cut, whose $\omega=0$ structure leads to the power law decay
\cite{Leaver}.
In Fig.~\ref{fig:late-time} the overlap of those different qualitative
behaviors is demonstrated from the viewpoint of future null infinity
(the initial data is a pulse of compact support located near the
horizon).

For the given \sch potential, the power of the late-time tail is
characteristic of generic compact support initial data, and depends
only on the multipole index $l$. At the time-like and null infinities,
the tail dependence on time is shown \cite{Price,Suen} to be
respectively $t^{-2l-3}$ and $u^{-l-2}$. Verifying this behavior
constitutes a good accuracy and stability test of the CCM algorithm,
as it follows the physical behavior of the system over a large
amplitude range, over long periods of time.  

The null infinity signal is read at $x=1$, i.e., the compactified
location of null infinity. The time-like infinity signal is read at
the matching boundary.  In Fig.~\ref{fig:null-power} the focus is on
the final power law behavior at null infinity. The run parameters are
a grid of 1500 points in each of the regions, a matching radius at
$r=20 M$, and an initial pulse with compact support between $\r \in
(-15,15)$.  In Fig.~\ref{fig:time-power} the focus is on the final
power law behavior at time-like infinity, with the same run parameters
as Fig.~\ref{fig:null-power}.

The late time power law tails at null infinity for various multipoles
($l=0,1,2$) are measured to be $-1.89, -2.91$ and $ -3.9$, in
agreement with the expected values of $-2, -3, -4$.  Similarly, the
late time power law tails at time-like infinity for the multipoles
$l=0,1$, are measured to be $-3.04$ and $-4.98$ respectively, which
should be compared to the theoretically expected values of $-3$ and
$-5$ respectively. To increase the range of multipole values whose
decay is observable, the grid resolution must be increased.

The monitoring of tails at both timelike and null infinity, with a
modest number or radial grid points, illustrates vividly the physical
and algorithmic economy achieved with the CCM approach. The late time
behavior of the field is directly related to the asymptotic fall-of of
the potential. This fall-off is captured easily in the compactified
description of the spacetime.

\section{Conclusions}\label{sec:conclusions}

The results presented in this paper accompany the discussion of GLPW
and complete the investigation of the CCM framework for the model
problem of a spherically symmetric scalar field.  These results lend
further support to the idea that a Cauchy + characteristic approach
for solving initial value problems is a potentially useful approach to
numerical relativity. The completion of the computational domain with
a null foliation extending to infinity eliminates the problem of
backscaterred waves in a natural way, even for arbitrary long evolution
times. It hence provides a significant increase in the dynamic range
of standard Cauchy codes.  There is, however, considerable amount of
work to be carried out before one can reach a definite answer as to
whether CCM approaches are computationally efficient in three
dimensional problems.

\section{Acknowledgments}

This work was supported by NSF grants PHY/ASC-9318152 (ARPA
supplemented), PHY-9357219 (P.L. NYI), and PHY-9309834. We thank
R.~Matzner, R.~Price, J.~Pullin for comments and helpful discussions.
P.~P. gladly acknowledges long discussions with J.~Winicour, R.~Gomez
and N.~Bishop on all aspects of the CCM problem.  Special thanks to
W.~M.~Suen and collaborators for making available to us details of their
numerical runs.

\newpage

\begin{figure}
\caption{Energy conservation. The upper plot displays the energy
content inside the world-tube as a function of time (diamonds) along with
the integrated flux of radiation across the world-tube (circles). The
lower plot demonstrates the convergence of the conserved
total energy as a function of grid spacing. The convergence rate
is measured as 1.995}
\label{fig:energy-con}
\end{figure}

\begin{figure}
\caption{Space-time evolution of a scalar field in the case of a
moving boundary. An oscillating source is located at $r=20$. The
original position of the boundary is at $r=50$ and it subsequently
moves inwards. To the left of the boundary (small $r$) the field
values are obtained from the Cauchy domain. To the right of
the boundary (large $r$) the values are given by the null code.
The flattening of the evolution profile to the right of the boundary
is the typical manifestation of outgoing radiation in retarded time
coordinates.}
\label{fig:moving}
\end{figure}

\begin{figure}
\caption{The scalar field at null infinity for initial data of 
compact support. The polynomial pulse is seen to register first
(at finite retarded time) and is then followed by damped 
oscillations. After the amplitude of the normal modes decays
below the long range tail signal, the latter becomes the 
dominant feature.}
\label{fig:late-time}
\end{figure}

\begin{figure}
\caption{Late time power law tails at null infinity. The measured
exponents for $l=0,1,2$ are $-1.89, -2.91$ and $ -3.9$ respectively,
which compare well with the analytic values of $-2,-3,-4$.}
\label{fig:null-power}
\end{figure}

\begin{figure}
\caption{Late time power law tails at time-like infinity. The measured
exponents for $l=0,1$ are $-3.04$ and $-4.98$ respectively. The
predicted values are ($-3, -5$)}
\label{fig:time-power}
\end{figure}

\end{document}